\documentclass[sigconf,10pt]{acmart}
\usepackage{hyphenat}

\begin{document}

\title[TryLinks]{TryLinks: An interactive tutorial system for a cross-tier Web
  programming language}

\author{Junao Wu}
\affiliation{%
  \institution{ }
}
\email{wujunao19950715@gmail.com}
\author{Arek Mikolajczak}
\affiliation{%
  \institution{ }
}
\email{arek.mikolajczak19@gmail.com}

\author{James Cheney}
\affiliation{%
  \institution{LFCS, University of Edinburgh }
}
\email{jcheney@inf.ed.ac.uk}

\begin{abstract}
Links is a web programming language under development in Edinburgh aimed at simplifying web development. Conventional multi-tier applications involve programming in several languages for different layers, and the mismatches between these layers and abstractions need to be handled by the programmer, which can lead to costly errors or security vulnerabilities. In Links, programs combine all of the code of a web application in a single program, and the implementation generates appropriate JavaScript and HTML for the client, and SQL queries for the database.

However, installing and using Links is non-trivial, making it difficult for new programmers to get started and learn about Links. This paper reports on a Web-based ``TryLinks" system, allowing anyone to experiment with Links without going through the installation process. TryLinks was designed with two major functionalities: an interactive Links shell that teaches the basic syntax of Links and acts as a playground, as well as a short tutorial series on how Links is used in practical web development. Tutorials can also be created or modified by administrators.
We present the design and implementation of TryLinks, and conclude with discussion of lessons learned from this project and remaining challenges for Web-based tutorials for Web programming languages.
\end{abstract}

\copyrightyear{2019}
\acmYear{2019}
\setcopyright{acmlicensed}
\acmConference[Programming '19]{Companion of the 3rd International Conference on Art, Science, and Engineering of Programming}{April 1--4, 2019}{Genova, Italy}
\acmBooktitle{Companion of the 3rd International Conference on Art, Science, and Engineering of Programming (Programming '19), April 1--4, 2019, Genova, Italy}
\acmPrice{15.00}
\acmDOI{10.1145/3328433.3328450} 
\acmISBN{978-1-4503-6257-3/19/04}

\maketitle

\section{Introduction}

Web applications are distributed programs.  Typically, computation is spread across the server, Web client (browser) and database in the \textit{three-tier model}. Each tier has a dedicated responsibility and an established protocol to communicate with other tiers.
The three-tier model separates concerns and allows for distribution of user-specific computation (for example for UI events) to the client.  In doing so it makes developing the applications more difficult, since each tier uses a different set of languages and technologies. mismatches between these layers and abstractions need to be handled by the programmer, and this can lead to costly errors and security vulnerabilities.

The Links programming language~\cite{cooper_lindley_wadler_yallop_2007} aims to address these difficulties, by using one language and uniting the 3 tiers. In Links, the web application is a single program (whose type-safety can be checked once and for all at a high level) and the implementation takes care of generating appropriate code for the layers comprising the web application.

However, installing and using Links for web or database programming is non-trivial, and involves not only installing Links itself but installing or configuring a database server, making it difficult for new programmers to learn enough about Links to decide whether it meets their needs. Also, Links has its own syntax, which is similar to some programming languages but still takes a certain period of time to get used to and master. For any new developer who decides to use Links, currently the ramp up time is rather long, with very little help during the process.

Interactive Web-based tutorials or read-eval-print loops are widely available for most major programming languages (e.g. codecademy), but to the best of our knowledge there are no such systems available for Web programming languages themselves.  We think it is interesting to develop some, firstly because doing so should increase the accessibility of such languages, and secondly because doing so may highlight challenges for Web programming languages themselves.  In particular, one might hope that a given Web programming language would be well-suited to developing an interactive Web tutorial application for itself.

In this paper we present a web-based ``TryLinks" tutorial system, which provides an interactive read-eval-print loop and tutorial lessons similar to those on the Links project website.  Although TryLinks was developed with Links in mind, TryLinks itself is implemented using conventional Web programming frameworks (Node.js, Angular Dart, Express, WebSockets etc.) Thus, a similar approach should work to support other Web programming languages.

In the rest of this paper, we give an overview of the capabilities of TryLinks, and then present some details of the implementation, focusing on the particular challenges faced in an online tutorial system that can build Web applications.  We conclude with a discussion of the strengths and weaknesses of TryLinks, and directions for future work.

\section{Demonstration}

TryLinks provides three basic capabilities:
\begin{itemize}
  \item An interactive mode providing a Links \emph{Read-Eval-Print Loop} that runs in
    a web browser.  
  \item A set of tutorials that cover the basics of Web
    programming in Links.
\item An administrator mode that allows new tutorials to be created or edited.
\end{itemize}

In this section we highlight the main features of these components of
TryLinks.  
\begin{figure}[tb]
  \centering
\includegraphics[width=\linewidth]{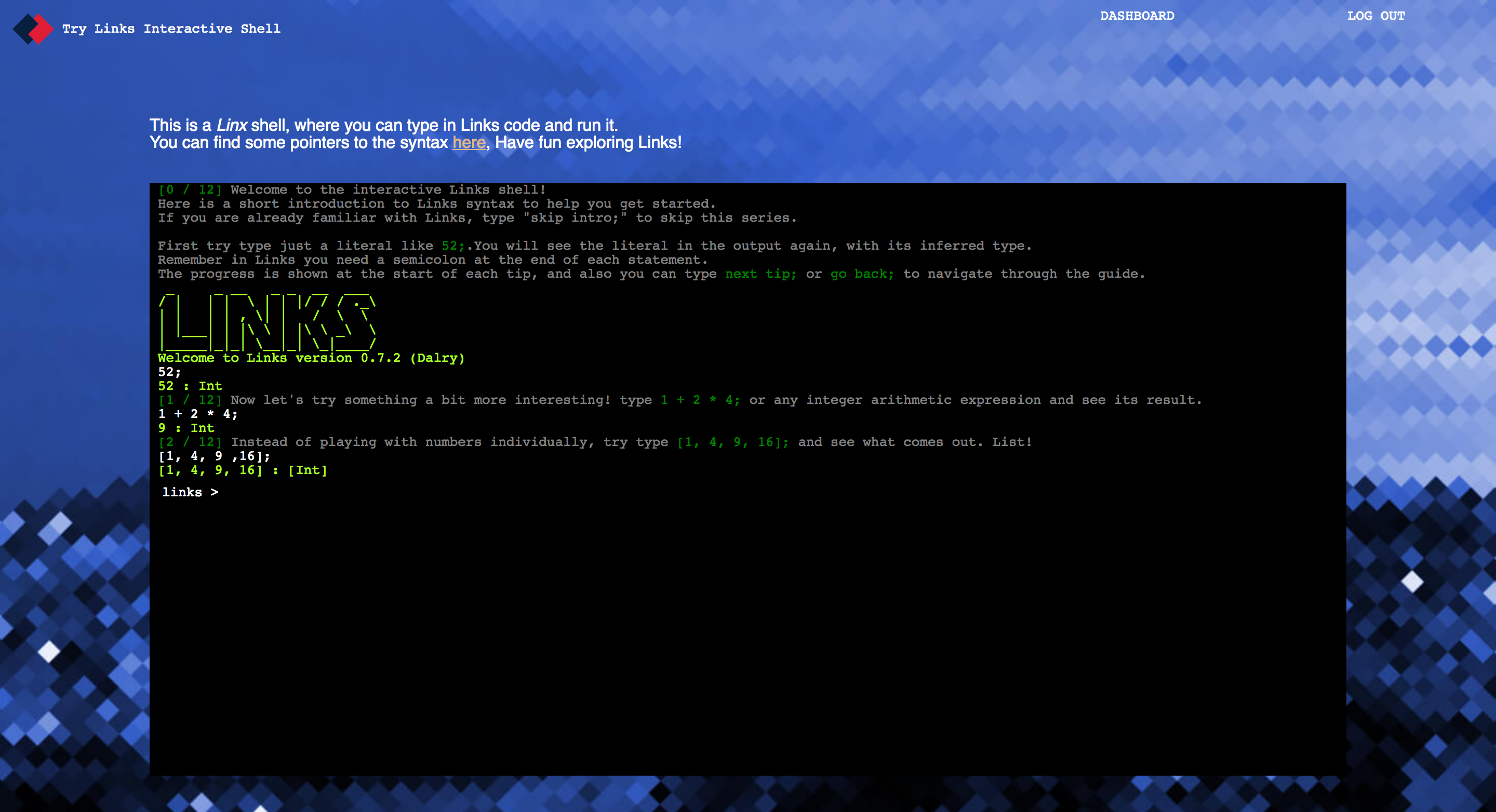}
  \caption{TryLinks read-eval-print loop}
  \label{fig:repl}
\end{figure}
\begin{figure}[tb]
  \centering
\includegraphics[width=\linewidth]{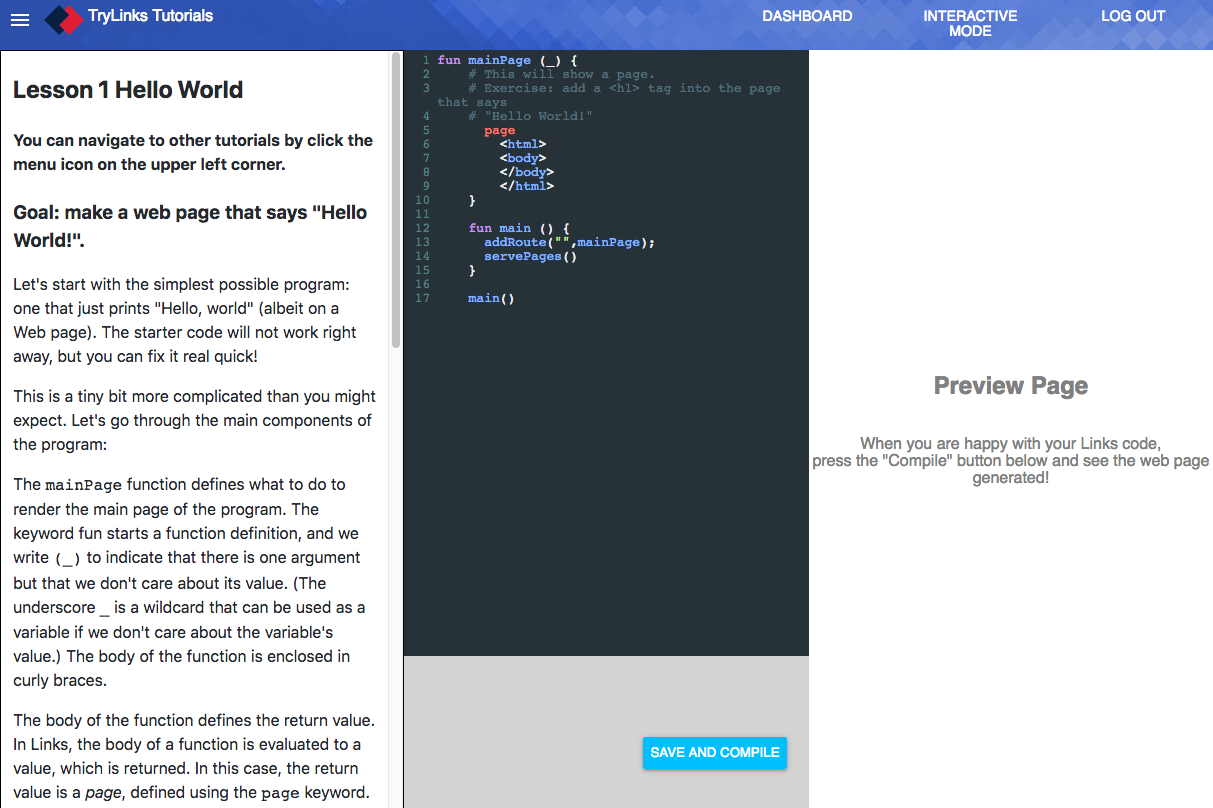}
  \caption{TryLinks tutorial}
  \label{fig:tutorial}
\end{figure} 

The REPL page (Fig.~\ref{fig:repl}) provides a connection to a Links REPL.  Links expressions can be entered in the standard way, and there is a sequence of hints suggesting different expressions the user may enter, starting with basic arithmetic, lists, pattern-matching, anonymous functions, and comprehensions.  The REPL supports the full Links language, although some capabilities are disabled for security reasons, and web applications cannot be run in this mode.

The tutorial page (Fig.~\ref{fig:tutorial}) consists of three frames: a problem description, a Links code editor, and a pane showing the running Links program.  The tutorials introduce the main concepts needed to create simple Links web applications:  creating static pages, simple forms and formlets, a simple TODO list, querying from a database, using queries and updates to create a persistent TODO list.  Finally the system also allows administrators to create or modify tutorial programs and descriptions.

The reader is encouraged to try the system
out using the dummy user account \texttt{testt} with password 12345.
TryLinks is available at \url{http://examples.links-lang.org/} and the
source code is available on GitHub from
\url{https://github.com/links-lang/}.  A demo video is available here: \url{https://tinyurl.com/y7fz44b6}.

\section{Implementation}

Although TryLinks is mostly a conventional three-tier Web
application, Links is unfortunately not yet a suitable implementation
language for TryLinks.  Instead, TryLinks is
implemented using a combination of standard Web technologies and is
subject to all of the usual complications of multi-language,
multi-tier development.  Links executables are run
as black-boxes and connected to the Web client using WebSockets.  In
this section we outline the current implementation strategy.

\subsection{TryLinks Server}
\begin{figure}[tb]
\centering
\includegraphics[width=\linewidth]{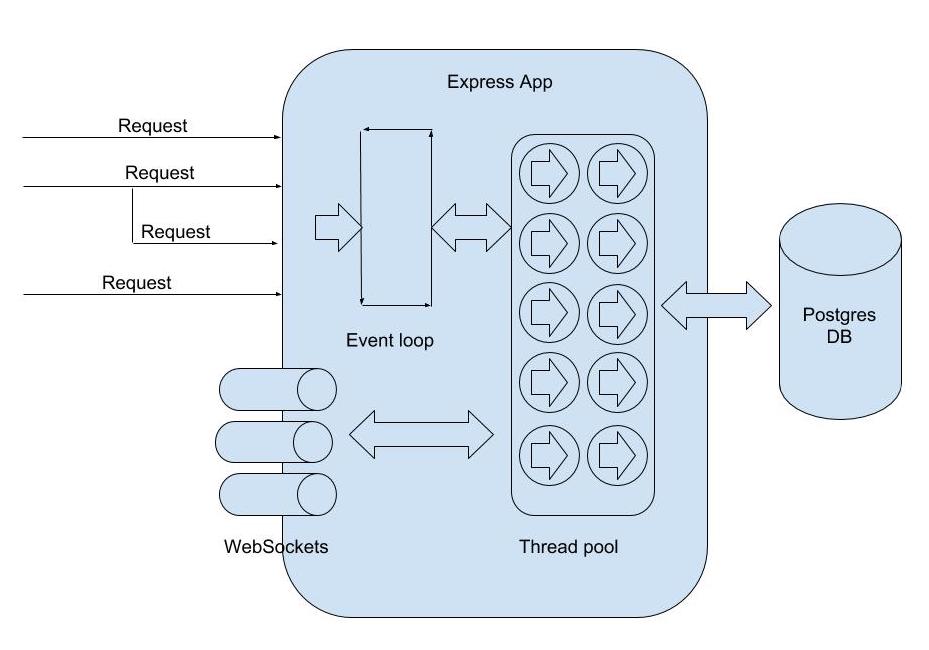}
\caption{TryLinks backend structure based on the \textit{Express} framework}
\label{fig:express_architecture}
\end{figure}
The backend was implemented in Node.JS using the Express framework.  
Figure \ref{fig:express_architecture} shows the Express backend structure, tuned for TryLinks. The Express application can be divided into two major components: the event loop and asynchronous thread pool. The event loop is a single thread that acts as a handler for incoming requests, and delegates each request to a thread in the thread pool. When a request is handled and a corresponding response is ready, the event loop picks it up and returns it to the sender. Using this structure promises the application is always ready to receive new requests, and handle heavy traffic with high level of concurrency.

What is different for TryLinks is the interactions with the threads
in the thread pool. Apart from the usual interactions with the main
event loop and with the database, the individual threads could also
create, close, and interact with WebSockets, which can be connected by
a client. 

In addition to Express, TryLinks also leverages the advantage of NPM,
and made use of many third party packages and libraries. Standard
packages for password processing, session management, database
interaction, spawning child processes were used.  We provide more
details of TryLinks's use of WebSockets because there were a number of
complications that needed to be resolved.

\paragraph{WebSocket}
The traditional HTTP communication between the client and the server is uni-directional. The client can only request and the server can only respond. This makes it difficult for the server to initiate a message to a client. Unlike HTTP, WebSocket provides ``full-duplex communication" \cite{websocket_definition}. This means both the server and the client can read and write on an established connection.

\begin{figure}[tb]
\centering
\includegraphics[width=\linewidth]{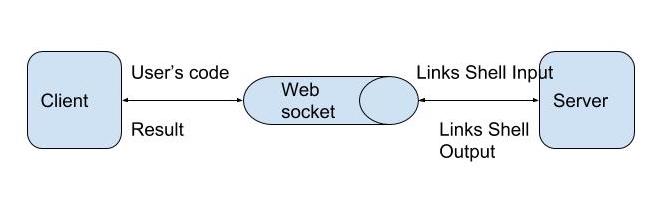}
\caption{WebSocket's role in connecting client and server for Links process}
\label{fig:websocket_links}
\end{figure}

WebSocket was essential to TryLinks, since the server and the client must send messages back and forth for the Links processes to be dynamic and interactive. Figure \ref{fig:websocket_links} shows how WebSocket facilitated the redirection of input and output, on both the client and server.

There are many WebSocket modules available in NPM; \textit{Socket.IO} \cite{socket.io} was used in TryLinks. It was easy to set up the connection and register callbacks for built-in and custom events, also \textit{Socket.IO} supported ``namespaces" for maintaining one WebSocket channel for each client. However, during development we ran across a bug in the Dart Socket.IO client. After contacting the author, the problem was solved.

\begin{figure}[tb]
\centering
\includegraphics[width=\linewidth]{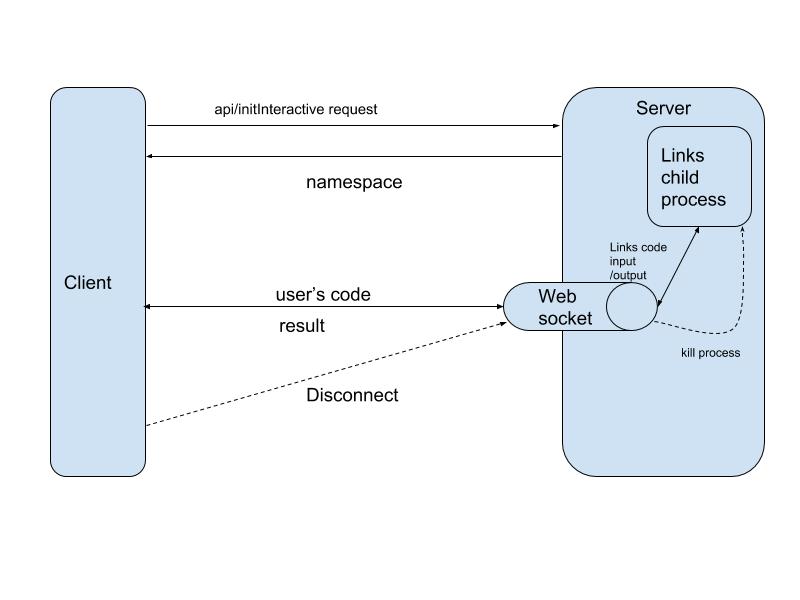}
\caption{TryLinks REPL implementation pipeline}
\label{fig:REPL_pipeline}
\end{figure}

\paragraph{REPL Implementation}
Using the Node \texttt{child\char`_process} module and WebSocket, the Links REPL shell was almost trivially implemented. Figure \ref{fig:REPL_pipeline} shows the basic pipeline of the REPL shell functionality. First of all, the server creates a new WebSocket channel, with the username as namespace, when the \texttt{api/initInteractive} HTTP endpoint is called, after checking for a valid session. In addition, the \texttt{api/init\hyp{}Interactive} HTTP endpoint also spawns a new Links child process, and redirects its input and output port to the WebSocket. Finally it responds with the namespace for the client to connect to.

The client then connects to the correct WebSocket. Upon successful connection a welcome message from Links is displayed, which informs the user that the REPL is ready. User inputs are sanitized according to Links syntax \cite{links_syntax} to ensure security. In a local Links shell, one could adjust various configuration settings by using the \texttt{@set} command. In TryLinks REPL shell this is disabled, and a special error message is shown when \texttt{@set} was encountered.

When the client disconnects from the WebSocket, usually by leaving the page, a built-in \texttt{disconnect} event is sent to the server. The server then disconnects the Links child process from the WebSocket, and sends a signal to terminate said process. 

\begin{figure}[tb]
\centering
\includegraphics[width=\linewidth]{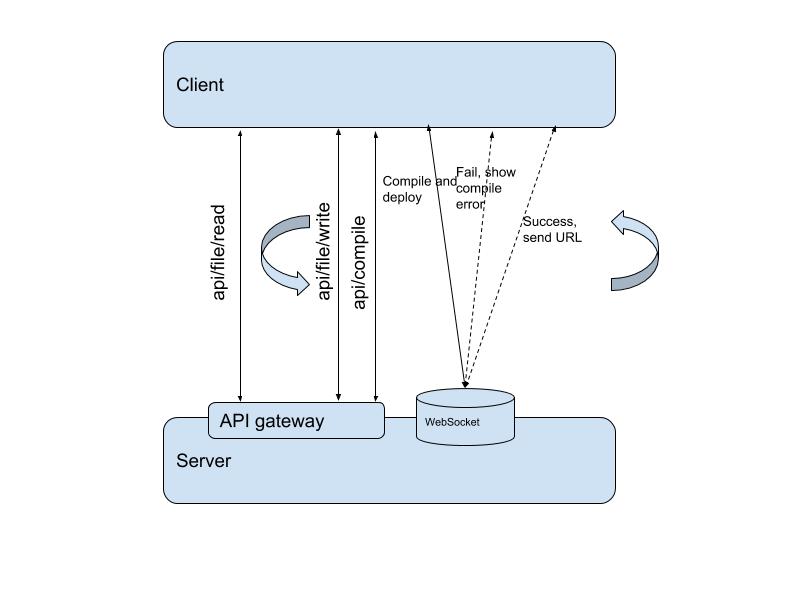}
\caption{TryLinks Links program compilation and deployment pipeline}
\label{fig:compile_pipeline}
\end{figure}

\paragraph{Compilation and Deployment Pipeline}
The Compilation and Deployment Pipeline refers to the process where the user requests to compile their custom Links program, which serves a web page. Upon successful compilation, the compiled web page is shown to the user. This functionality was also implemented with \texttt{child\char`_process} module and WebSockets. Figure \ref{fig:compile_pipeline} shows the implementation pipeline of this process.

A common development pattern for web developers starts with editing the source code, then saving and compiling the code, observing the rendered web page, and back to editing. In TryLinks, this pattern drove the design of the compile and deploy pipeline. First of all, the stored source code is pulled from the database. The user makes changes to the source code, then requests to save and compile.

If the compilation is successful, the port to access the web page is
sent back to the user. Otherwise, compile errors are shown
instead. The user can make further changes and so on. The use of WebSockets was to catch compile errors, and enable
re-compile functionality.

\subsection{TryLinks Client}

The TryLinks frontend was designed as a simple MPA (Multi-Page
Application), with 3 main pages: \textit{dashboard},
\textit{interactive shell}, and \textit{tutorial} page. This decision
was inspired mostly by Codecademy. To give strong consistency with the
official Links website, TryLinks adopted the same Links logo and same
basic background image. In addition, material design
\cite{material_design} components were used extensively across the
frontend to enhance user experience.

\paragraph{Angular Dart}
Angular Dart \cite{angular_dart} is a component based web development
framework, loosely following the MVC (Model-View-Controller) pattern.
Each Angular Dart \textit{component} controls a part of screen called
a \textit{view}, which is defined by the component's binding
\textit{template}. Components nest to form the entire application. To
include child components, the parent component must declare them in
its \textit{metadata}, which also contains more information about the
component.

\textit{Service} is a broad category encompassing any value, function, or feature that the application needs. A service is typically a class with a narrow, well-defined purpose. Components utilize services by ``injecting" the services into themselves. This is called \textit{Dependency Injection}. Dependency injection ``separates the creation of a client's dependencies from the client's behavior, which allows program designs to be loosely coupled" \cite{seemann_2010} and to ``follow the dependency inversion and single responsibility principles" \cite{DBLP:journals/jot/SchwarzLN12}.

\begin{figure}[t]
\centering
\includegraphics[width=\linewidth]{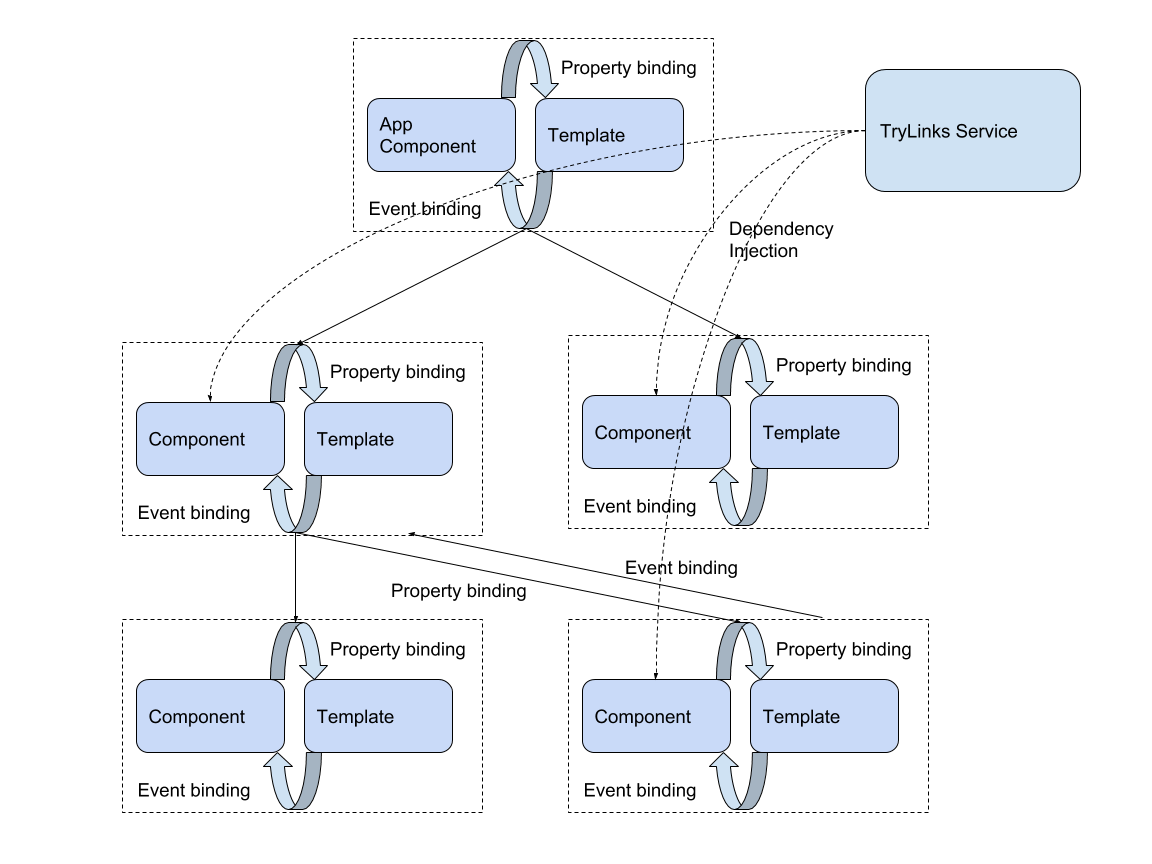}
\caption{TryLinks frontend architecture}
\label{fig:TryLinks_arch}
\end{figure}

TryLinks was implemented following the best practices of Angular Dart. Each page was well broken down into components; data and event binding were used extensively both intra- and inter-components; a custom service was built to interact with the backend as well as manage session data. Figure \ref{fig:TryLinks_arch} shows the architecture of the frontend of TryLinks, slightly tweaked from the standard architecture.

\paragraph{Interactions with Backend}
As mentioned before, a custom \texttt{TryLinksService} was built for
interactions with backend. It contained all HTTP calls and a few
helper functions that supplied as application global variables,
including a configurable parameter for the service base URL. This was
helpful in that the URL could be customized without changing the
source, which accelerated the deployment process and reduced the
possibility of using the wrong URL.

\section{Discussion}

\subsection{Performance}
Some effort has been made to ensure TryLinks has acceptable
performance.  Page loading time for TryLinks and several other
language tutorial sites  have been measured (with
low load) using Pingdom\cite{pingdom}, as follows: 

\begin{tabular}{l|c|c}
System & Page Size & Load Time\\\hline
  TryLinks & 1.0 MB & 1.06 s\\
TryHaskell & 125.7 kB & 824 ms\\
repl.it & 3.5MB & 2.08s\\
W3School & 453.6 kB & 1.09 s\\
Codecademy & 2.4 MB  & 2.60 s\\
JsFiddle & 646.7 kB & 1.70 s\\
Python online shell & 727.3 kB & 2.96 s
\end{tabular}

These times measure the front page only, however.  Subsequent pages, such as
the REPL and Links tutorial pages, have a longer start-up time or
delays when a program is recompiled and the Links executable is stopped
and restarted.

During the evaluation with the Links team, it was observed that
sometimes the REPL page took about 2 seconds to be loaded and
functional. Further investigation showed that this was caused by the
Links child process starting up. Although this was technically out of
scope of the project, it did affect usability of TryLinks. This may
serve as a motivation for Links to refactor the core library to speed
up loading time, or to implement a dynamic re-loading functionality. 

\subsection{User feedback}
TryLinks was designed with the goal of helping developers to learn the Links programming language; as a result reviewing this goal was the main focus of evaluation. In this section the evaluation methodology is presented, and the main findings are summarized.

The TryLinks evaluation session was designed as follows. Each user filled out a pre-evaluation questionnaire, which consisted of mostly background questions. Then they launched TryLinks in the browser, and used the software for an hour. At the end, a post-evaluation questionnaire was filled out, which tested the user's understanding of the Links language, and also included user experience feedback.

The process of evaluation was written into a \textit{README}. In it some more detailed description of TryLinks was also given to aid the user.

In terms of evaluation group demographics, 9 people with Web or functional programming experience but no previous
Links experience were selected. For this group the main evaluation
focus was on if TryLinks helped them to learn the Links language. In
addition, 2 people from the Links team also participated in the
evaluation. For this group the focus was on the organization and
presentation of TryLinks, in other words did TryLinks present Links in
a way which is easy to learn. Note that the second group's feedback on
learning the Links language is not relevant to the main goal and
therefore not included in the findings. However, their critical
evaluations are discussed later in Section \ref{sec:limitations}.

The full experimental results are reported in ~\cite{trylinks-report};
we focus on the main result of interest, namely whether the
participants were able to use TryLinks to complete tutorials.

\begin{figure}[tb]
  \centering
  \includegraphics[width=\linewidth]{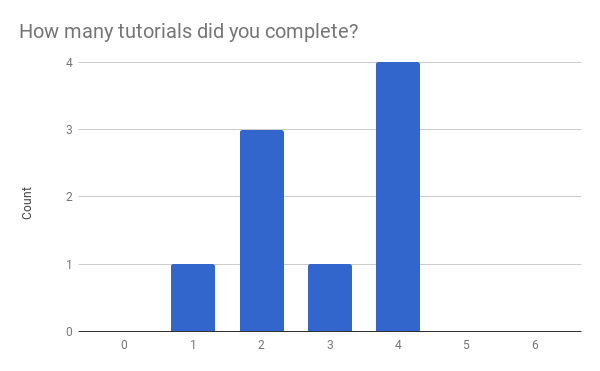}
  \caption{Response of how many tutorials the participants completed under an hour}
  \label{fig:chart_tutorial_complete}
\end{figure}

Figure \ref{fig:chart_tutorial_complete} shows the responses of this question. Following up from the last point, no participants completed more than 4 tutorials under an hour, and most of them reported that it was hard to progress through the tutorials. This is so partly because the tutorials themselves were not constructed in the best way. One participant commented: ``It is a bit abstract, between learning the language Links and the usage in web programming." The transition from using Links in the command line to using it for web programming is so sharp that some users lose the narrative.

However, this was actually a quite positive result, since before TryLinks it was impossible for non-experts to get Links running in an hour, let alone using it in any tutorial. In this regard, TryLinks massively shortened the installation time, and in turn lowered the difficulty of getting started with Links.

\subsection{Limitations}\label{sec:limitations}

During the development and evaluation of TryLinks, a number of bugs
and user interface problems were fixed.  We discuss issues that have
not been easy to fix, due to limitations in Links, in the conventional
web programming frameworks we are using, or both.

\paragraph{Lack of evaluation system}
TryLinks does not have a built-in evaluation system. Instead, users
need to read the feedback in the introduction guide on the REPL page
to determine whether their responses are correct.  Likewise, during
the tutorials it is sometimes not clear what the goal is and this is
not automatically checked (aside from checking whether the code
compiles).  Our users would have appreciated a built-in evaluation
system, so that they feel more certain that they have coded
correctly. While testing that REPL entries match the expected script
is conceptually straightforward, it would require parsing Links
results and/or modifying the underlying REPL code.  Testing that the
submitted solution to an interactive, Web tutorial example meets a
specification would likewise require approximating Links program
equivalence somehow, which is also nontrivial.

\paragraph{Extensibility and admin mode}
Another limitation we observed ourselves is that it is not ideal to
have a fixed set of tutorials.  Adding a tutorial or its supporting
documentation then requires manually adding entries to the underlying
database.  As mentioned earlier, we have experimented with an ``admin mode'', in which
certain users have the privilege to add new tutorials or edit the
initial state of existing
ones. However, this is still somewhat limited: there is a single list of tutorials, and new tutorials can only be added at the end.  Allowing for multiple groups of tutorials, and for rearrangement of the order of the tutorial list, would make this feature more useful.

\paragraph{Login requirement}
Currently, TryLinks requires the user to create an account and log in
before accessing the REPL or tutorial settings.  This is because the
Links system runs on the server so starting a new Links process for
each access to the REPL or tutorial page is infeasible.  Requiring
users to log in enables us to limit the number of processes created
per user, so that one user cannot deny access to others or crash the
system.  It also means that we can save users' partial progress and
they can return later.  However, many other language learning sites provide at least
a basic REPL or tutorial that can be accessed without any login
credentials.  This is accomplished by providing a stripped-down REPL
implementation that can run on the client side.  For Links, this has
not been implemented but tools such as \verb|js_of_ocaml| might
facilitate this process.

\section{Conclusions}

TryLinks is a functional online platform for everyone to try out the Links programming language. Initially planned as an experimental project, TryLinks has received positive feedback from its target audience.

Throughout the development process, several obstacles were encountered, and finally fixed. One of them was using WebSocket with child processes, especially for the ``compile and deploy" functionality. It was confusing, at first, to have to register callbacks to the child process input and output ports when the WebSocket first connects, even though the process itself might not have been booted up and functional. Furthermore, in the case of compiling Links programs, when the child process just starts and does not give any output, it is impossible to distinguish whether the process is still booting up, or the compilation has been successful, due to Links core library implementation. This issue forced a complete rewrite of the pipeline, which integrated WebSockets and carefully mapped the sequences of outputs to each scenario.

Another great difficulty was to implement the tutorial page. Being the most complicated page in the whole site, it utilized many technologies and modules. Creating the layout was difficult due to the peculiarities of these modules. In addition, coordinating these technologies presented some difficulties as well.

Last but not least, the first version of TryLinks was not very user
friendly, and the initial alpha test reported many problems with the
site's usability. Also bugs were found throughout the system, from
trivial ones such as spelling mistakes in the tutorial descriptions,
to critical ones such as the rendering port overlapping issue. It took
some time to process the comments and rethink about the site.
Fortunately, most of these problems were fixed before the evaluations
and the final version of TryLinks turned out to be much better.

TryLinks shows that it is possible, using a combination of conventional Web technology and libraries, to provide a Web-based interface for Web programming.  However, TryLinks's design also highlights shortcomings in both conventional Web development and cross-tier languages such as Links.  In the future, it would be interesting to extend Links-style cross-tier programming to enable writing a Web tutorial for Links in Links itself.

\paragraph*{Acknowledgments}
  This work was supported by ERC Consolidator Grant Skye (grant number
  \grantnum{ERC}{682315}).

%
\bibliographystyle{ACM-Reference-Format}
\bibliography{references}

\end{document}